\documentclass[conference,final,twocolumn,10pt]{IEEEtran}
\IEEEoverridecommandlockouts
\newtheorem{lemma}{Lemma}
\newtheorem{theorem}{Theorem}
\newtheorem{corollary}{Corollary}

\usepackage{graphicx}
\usepackage{epstopdf}
\usepackage{stfloats}
\usepackage{subfig}
\usepackage{cite}
\usepackage{amsmath}
\usepackage{caption}
\usepackage{amsfonts,amssymb}
\usepackage[aboveskip=10pt,belowskip=-15pt,font=footnotesize]{caption}
\usepackage[CJKbookmarks=true,bookmarksnumbered=true,bookmarksopen=true]{hyperref}
\usepackage{setspace}
\usepackage{color}

\let\oldcite\cite
\renewcommand{\cite}[1]{\mbox{\oldcite{#1}}}

\ifCLASSINFOpdf
\else
\fi

\begin{document}
	
	\title{\huge Performance Evaluation for Subarray-based Reconfigurable Intelligent Surface-Aided Wireless Communication Systems}
	
	\vspace {-2em}
	\author{
	\IEEEauthorblockN{
		Xinyi~Yang, 
		Weicong~Chen, 
		Xiao~Li, 
		and~Shi~Jin}\\	
			\vspace {-1.2em}

    \vspace {-2em}

	\thanks{{Xinyi Yang, Weicong Chen, Xiao Li, and Shi Jin are with the National Mobile Communications
			Research Laboratory, Southeast University, Nanjing 210096, China (e-mail: yangxinyi@seu.edu.cn; cwc@seu.edu.cn; li\_xiao@seu.edu.cn; jinshi@seu.edu.cn).} }	
		
		
	}

	\maketitle

	
	\begin{abstract}
		Reconfigurable intelligent surfaces (RISs) have received extensive concern to improve the performance of wireless communication systems. 
		In this paper, a subarray-based scheme is investigated in terms of its effects on ergodic spectral efficiency (SE) and energy efficiency (EE) in RIS-assisted systems.
		In this scheme, the adjacent elements divided into a subarray are controlled by one signal and share the same reflection coefficient. 
		An upper bound of ergodic SE is derived and an optimal phase shift design is proposed for the subarray-based RIS. Based on the upper bound and optimal design, we obtain the maximum of the upper bound.
		In particular, we analytically evaluate the effect of the subarray-based RIS on EE since it reduces SE and power consumption simultaneously.
		Numerical results verify the tightness of the upper bound, demonstrate the effectiveness of the optimal phase shift design for the subarray-based RIS, and reveal the effects of the subarray-based scheme on SE and EE.
	\end{abstract}
	

	\IEEEpeerreviewmaketitle
	\section{Introduction}
	Reconfigurable intelligent surface (RIS) has emerged as a rising popular technology thanks to its ability to realize a smart radio environment that is customizable\cite{2019arXiv190308925D,chen2021channel}. A typical RIS shows as a planar surface comprising numbers of reflecting elements. By imposing phase shift on the impinging signals to collectively reflect them in the desired direction, the RIS can enhance the received signal and expand the coverage area\cite{RN13}. However, this promising technology still faces several challenges from the practical point of view, especially when large amounts of elements are employed to compensate for the even severer power loss in the reflecting path\cite{RN14}. The large quantities of RIS elements directly contribute to the unfavorable complexity of the reflection mode setting as well as optimization due to the numerous reflection coefficients that need to be dealt with,
    and the multiplication of the number of channel state information (CSI) coefficients between the transmitter and receiver introduced by RIS makes the cascade channel estimation quite costly and challenging\cite{RN16,9053695}. Moreover, as the element number increases, the power consumption of RIS increases as well\cite{2022arXiv221100323W}.
	
	To cope with the problems above, subarray-based schemes are adopted to reduce high complexity and overhead. Reference\cite{RN15} utilized the subarray-based strategy to reduce the pilot overhead and computational complexity of channel estimation and reflection optimization, where the adjacent elements are grouped into a subarray to share the same reflection coefficient 
	and considered to possess a high channel correlation while the phase difference of signal due to their spacing is ignored. The training overhead reduction introduced by the grouping strategy in the orthogonal frequency division multiplexing (OFDM) system is also evaluated from the aspect of symbol durations\cite{RN16}. In addition, the advantages of subarray-based schemes in different communication scenarios are investigated to improve the achievable rate\cite{RN17}, and minimize power consumption by adjusting pilot length\cite{2023arXiv230316625E} and reducing transmit power\cite{RN19}. A sparse array of sub-surface deployment was proposed in\cite{9322466}, and the authors showed that the correlation of the effective channels for different users can be reduced when the distances of sub-surfaces are set properly. However, the effect of the power reduction advantages of the subarray-based schemes on performance in terms of energy efficiency (EE) has not been investigated sufficiently.
	
	In this letter, the impacts of the subarray-based RIS on spectral efficiency (SE) and EE are evaluated. Although the subarray-based RIS loses control degree of freedom, it requires less power consumption and is preferred when a certain degree of trade-off between SE and EE is necessary.
	An upper bound of ergodic spectral efficiency is first derived by exploiting statistical channel state information (CSI). Then, under the model of grouping adjacent RIS elements into subarrays with the same reflection coefficient, we obtain the optimal phase shift design to maximize the upper bound. The gap of ergodic SE and EE between the subarray-based scheme and conventional element-based scheme is comparatively analyzed on the basis of analytical expression. Simulation results confirm the tightness of the upper bound and the effectiveness of our deduced optimal phase shift design. Moreover, the results reveal how the ergodic SE and EE depend on the element numbers and the subarray size.
	\section{System Model}\label{section:System_model}
\begin{figure}[htbp]
		\centering
		\includegraphics[width=0.4\textwidth]{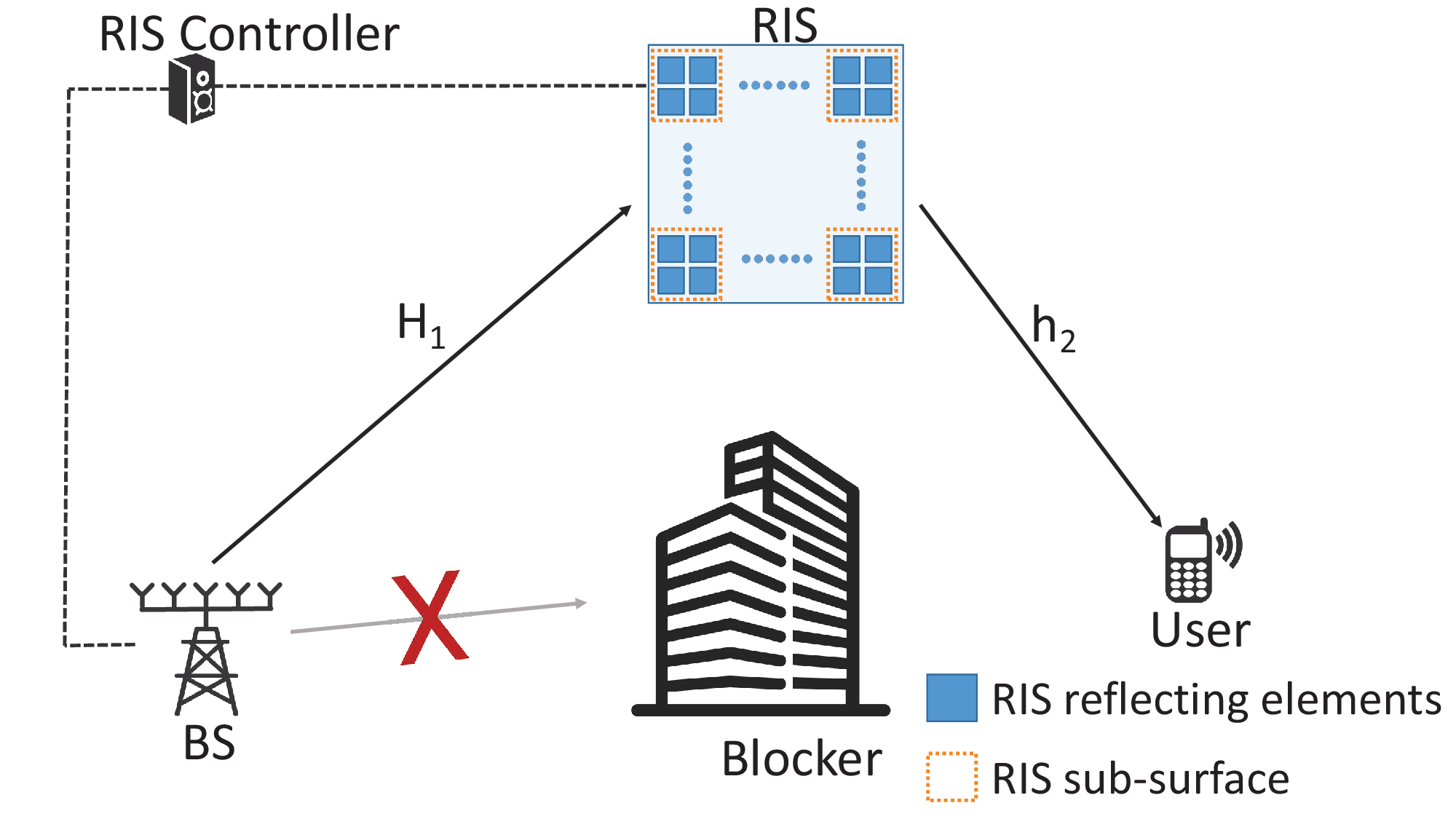}
		\caption{. The subarray-based RIS-assisted downlink communication system.}
		\label{fig:imageofsystem}
	\end{figure}

	The considered subarray-based RIS-assisted large-scale antenna system is illustrated in Fig. \ref{fig:imageofsystem}. A BS equipped with an $ M $-element uniform linear array (ULA) serves a single-antenna user. The line-of-sight (LoS) path between BS and the user is blocked. To address the blockage problem, a RIS is deployed between the BS and the user to assist the communication and arranged as a uniform planar array (UPA). A varactor-diode-based RIS is adopted to enable continuous phase shift adjustment. In the subarray-based RIS, $ N = {N_x} \times {N_y} $ passive reflecting elements are divided into $ Q = {Q_x} \times {Q_y} $ subarrays, each of which consists of $ L = {L_x} \times {L_y} = \frac{{{N_x}}}{{{Q_x}}} \times \frac{{{N_y}}}{{{Q_y}}} $ elements. To reduce power consumption, elements of the same subarray are controlled by one control circuit and share the same reflection coefficient.
	
	\subsection{Channel Model}  
	
	\vspace{-0.3cm}
	\vspace{+0.1cm}

	Due to a large amount of scattering in the wireless environment, the non-LoS (NLoS) component between BS and the user can be modeled as Rayleigh fading and denoted as $ \mathbf{g}\in \mathbb{C}^{1 \times M} $ where the elements of $ \mathbf{g} $ are i.i.d. and follow the complex Gaussian distribution of zero mean and unit variance. For the channels between BS and RIS and between RIS and the user, they are modeled as Rician fading. With Rician factor  ${K_1}$, the channel between BS and RIS can be expressed as
	\begin{equation}
		\mathbf {H}_1 = \sqrt{\frac{K_1}{K_1+1}}\mathbf {\bar H}_1 + \sqrt{\frac{1}{K_1+1}}\mathbf {\tilde H}_1,
		\label{H_1}
	\end{equation}
	where $ \mathbf{\bar H}_1\in \mathbb{C}^{N \times M} $ is the LoS component and $ \mathbf {\tilde H}_1\in \mathbb{C}^{N \times M} $ is the NLoS component.
	
	The elements of $\mathbf {\tilde H}_1 $ are i.i.d. in the complex Gaussian distribution with zero mean and unit variance. The LoS component is given by
	\begin{equation}
		{\bf{\bar H}}_1={\bf{ b}}^T\otimes[{\bf{a}}_L^H({\theta _{a_1}},{\varphi _{a_1}}){{\bf{a}}_M}({\theta _{d_1}})],
		\label{eq:H1bar}
	\end{equation} 
	where ${{\bf{ b}}} =[b_1,...,b_q,...,b_Q]\in {\mathbb{C}^{1 \times Q}} $. ${b_q}$ denotes the phase difference introduced by the position of the $q$-th subarray which is given by
	\begin{equation}
		{b_q} = {e^{ - j2\pi \frac{{{d_2}}}{\lambda }[\sin {\theta _{a_1}}  ({x_q} - 1) + \cos {\theta _{a_1}}\sin {\varphi _{a_1}}  ({y_q} - 1)]}},
	\end{equation}
	where $ {x_q} $ and $ {y_q} $ are the horizontal and vertical coordinates of the first element of the $q$-th RIS subarray, ${\theta _{a_1}}$ and ${\varphi _{a_1}}$ denote the elevation and azimuth angles of arrival (AoA) at RIS, and ${\theta _{d_1}}$ is the angle of departure (AoD) from the ULA at BS. The array steering vector ${{\bf{a}}_M}({\theta})\in {\mathbb{C}^{1 \times M}} $ for BS and ${\bf{a}}_L({\theta },{\varphi}) \in {\mathbb{C}^{1 \times L}}$ for RIS subarray can be expressed as 
	\begin{equation}
		{{\bf{a}}_M}({\theta}) = [1,{e^{j2\pi \frac{{{d_1}}}{\lambda }\sin {\theta}}},...,{e^{j2\pi \frac{{{d_1}}}{\lambda }\sin {\theta} (M - 1)}}],
	\end{equation}
	\begin{equation}
		\hspace{-0.37cm}
		\begin{aligned}
			{\bf{a}}_L({\theta},{\varphi}) = [1,{e^{j2\pi \frac{{{d_2}}}{\lambda }\sin {\theta}}},...,{e^{j2\pi \frac{{{d_2}}}{\lambda }\sin {\theta} ({L_x} - 1)}}]\\
			\otimes [1,{e^{j2\pi \frac{{{d_2}}}{\lambda }\sin {\varphi}\cos {\theta}}},...,{e^{j2\pi \frac{{{d_2}}}{\lambda }\sin {\varphi}\cos {\theta}  ({L_y} - 1)}}],
		\end{aligned}
	\end{equation}
	where ${d_1}$ is the antenna interval of BS, and ${d_2}$ is the vertical and horizontal spacing between RIS elements. Similarly, the channel between the RIS and the user is denoted as 
	\begin{equation}
		{{\bf{h}}_2} = \sqrt {\frac{{{K_2}}}{{{K_2} + 1}}} {{\bf{\bar h}}_2} + \sqrt {\frac{1}{{{K_2} + 1}}} {{\bf{\tilde h}}_2},
	\end{equation}
	where ${K_2}$ is the Rician factor and ${{\bf{\tilde h}}_2} \in {\mathbb{C}^{1 \times N}}$ is the NLoS component whose elements are i.i.d. complex Gaussian distributed with zero mean and unit variance. The LoS component ${{\bf{\bar h}}_2} \in {\mathbb{C}^{1 \times N}}$ is expressed as
	\begin{equation}
		{{\bf{\bar h}}_2} = {\bf{c}} \otimes {{\bf{a}}_L}({\theta _{d_2}},{\varphi _{d_2}}),
	\end{equation}
	where ${\bf{c}}=[c_1,...,c_q,...,c_Q] \in {\mathbb{C}^{1 \times Q}}$ with ${c_q}$ being the phase difference due to the position of the $q$-th subarray and can be given by
	\begin{equation}
		{c_q} = {e^{j2\pi \frac{{{d_2}}}{\lambda }[\sin {\theta _{d_2}}  ({x_q} - 1) + \cos {\theta _{d_2}}\sin {\varphi _{d_2}} ({y_q} - 1)]}}.
	\end{equation}
	
	\subsection{Power-consumption Model}
	The total power consumption of the system can be given by
	\begin{equation}
	{P_{{\rm{tot}}}} = \xi P + {P_{{\rm{BS}}}} + {P_{{\rm{UE}}}} + {P_{{\rm{RIS}}}}  \buildrel \Delta \over =  {P_{{\rm{rest}}}} + {P_{{\rm{RIS}}}},
	\label{P_tot}
    \end{equation}
	where $\xi$ is the power amplifier efficiency, $P$ is the transmission power, and ${P_{{\rm{BS}}}}$, ${P_{{\rm{UE}}}}$, and ${P_{{\rm{RIS}}}}$ denote the circuit power consumption at the BS, the user, and the RIS respectively.

	According to results in\cite{2022arXiv221100323W} and\cite{2023arXiv230300299W}, the power consumption of RIS is modeled as
	\begin{equation}
		{P_{{\rm{RIS}}}}{\rm{ = }}{P_{{\rm{dy}}}} + {P_{{\rm{st}}}}
		= {P_{{\rm{dy}}}} + {P_{{\rm{ct}}}} + {P_{{\rm{td}}}},
		\label{P_RIS}
	\end{equation}
	where ${P_{{\rm{dy}}}}$ is the dynamic power consumption and ${P_{{\rm{ct}}}}$ is the power consumption of the control board which can be regarded as a constant value. ${P_{{\rm{td}}}}$ is the total power consumption of the driving circuit at RIS which can be expressed as
	\begin{equation}
		{P_{{\rm{td}}}} = {N_{{\rm{d}}}} \cdot {P_{{\rm{d}}}},
		\label{P_total drive circuits}
	\end{equation}	
	where ${P_{{\rm{d}}}}$ is the power consumption of a single driving circuit, and the number of driving circuits ${N_{{\rm{d}}}}$ is proportional to the number of units $N$ and subarrays $Q$ respectively in the element-based and subarray-based RIS design.
	\vspace{-0.15cm}
	\subsection{Downlink Ergodic Spectral Efficiency and Energy Efficiency}
	In the considered scenarios, the received signal on the user side can be expressed as
	\begin{equation}
		r = \sqrt P ({{\bf{h}}_2}{\bf{\Phi }}{{\bf{H}}_1} + {\bf{g}}){{\bf{f}}^H}s + w,
	\end{equation}
	where $P$ corresponds to the transmit power at the BS. ${\bf{\Phi }} = {\rm{blkdiag}}\left\{ {{{\bf{\Phi }}_1},...,{{\bf{\Phi }}}_q,...,{{\bf{\Phi }}}_Q} \right\} \in {\mathbb{C}^{QL \times QL}}$ denote the phase shift matrix of RIS, where ${{\bf{\Phi }}_q} = {e^{j{\phi _q}}}  {\bf{I}}_{L\times L}$ 
	is the phase shift matrix of the $q$-th subarray, ${\phi _q} \in [0,2\pi )$ represents the phase shift shared by each element of the $q$-th subarray. ${\bf{f}} \in {\mathbb{C}^{1 \times M}}$ is the beamforming vector  satisfying ${\| {\bf{f}} \|^2} = 1$, $s$ is the transmitted data signal satisfying $\mathbb{E}\{ {{{| s |}^2}} \} = 1$, and $w$ is an additive white Gaussian noise with the elements independently drawn from ${{\cal CN}}{\rm{(0,}}\sigma _w^2)$, where $\sigma _w^2 = 1$ is the noise variance. 
	
	Denoting ${\bf{h}} = {{\bf{h}}_2}{\bf{\Phi }}{{\bf{H}}_1}$, we assume ${\bf{h}} + {\bf{g}}$ is known because it can be estimated at the BS\cite{9053695}.
	When the maximum ratio transmitting (MRT) is adopted at the BS, the beamforming vector ${\bf{f}}$ can be expressed as
	\begin{equation}
		{\bf{f}} = ({{{\bf{h}} + {\bf{g}}}})/{{\left\| {{\bf{h}} + {\bf{g}}} \right\|}}.
		\label{f}
	\end{equation}
	
	When the MRT is adopted, the ergodic SE can be given by
	\begin{equation}
		C = \mathbb{E}\left\{ {{{\log }_2}\left( {1 + ({P}/{{\sigma _w^2}}){{\left\| {{{\bf{h}}_2}{\bf{\Phi }}{{\bf{H}}_1} + {\bf{g}}} \right\|}^2}} \right)} \right\}.
		\label{SE_initial}
	\end{equation}
	Accordingly, ergodic energy efficiency can be denoted as
	\begin{equation}
		R = {{C}}/({{{P_{{\rm{rest}}}} + P_{{\rm{RIS}}}}}).
		\label{EE_second}
	\end{equation}
	
	Observing at (\ref{EE_second}), the ergodic EE of the system depends on its ergodic SE and the total system power consumption. Note that for the varactor-diode-based RIS whose dynamic power consumption can be neglected\cite{2022arXiv221100323W}, when the partitioned setting of RIS is fixed, i.e., when the number of control signals required by RIS is constant, the total power consumption of the system is independent of the phase setting. Then maximizing the ergodic EE is equivalent to maximizing the ergodic SE. Thus, we first study how to improve SE.
	\section{Ergodic Spectral Efficiency Analysis}\label{section:Analysis}
	In this section, a tight upper bound of the ergodic SE is derived. Based on the upper bound, we investigate the optimal design of ${\bf{\Phi }}$ to maximize the upper bound of ergodic SE. After that, the effects of Rician factors and $\eta $ are evaluated.
	\subsection{Upper Bound of Ergodic Spectral Efficiency}
	
	To gain a direct perception of the ergodic SE, we provide an upper bound of the ergodic SE as shown in \textit{Lemma 1}.
	\begin{lemma}
		In the considered RIS-assisted large-scale systems, the upper bound ergodic SE is approximated by
		\begin{equation}
			{C^{\rm{ub}}} = {\log _2}(1 + P({\gamma _1}{\left\| {{{{\bf{\bar h}}}_{2}}{\bf{\Phi }}{{{\bf{\bar H}}}_{1}}} \right\|^2} + {\gamma _2}MN + M)),
			\label{eq:SE_ub}
		\end{equation}
		where ${\gamma _1}$ and ${\gamma _2}$ are given by
		\begin{equation}
			{\gamma _1} = \frac{{{K_1}{K_2}}}{{({K_1} + 1)({K_2} + 1)}},{\gamma _2} = \frac{{{K_1} + {K_2} + 1}}{{({K_1} + 1)({K_2} + 1)}}.
			\label{gamma}
		\end{equation}
	\end{lemma}
	
	\begin{IEEEproof}
		According to Jensen's inequality, the ergodic SE can be upper-bounded by
		\begin{equation}
			C \le {\log _2}\left( {1 + P\mathbb{E}\left\{ {{{\left\| {{{\bf{h}}_2}{\bf{\Phi }}{{\bf{H}}_1} + {\bf{g}}} \right\|}^2}} \right\}} \right).
		\end{equation}	
		Following the analysis process in\cite{RN10}, we obtain
		\begin{equation}
			\mathbb{E}\left\{ {{{\left\| {{{\bf{h}}_2}{\bf{\Phi }}{{\bf{H}}_1} + {\bf{g}}} \right\|}^2}} \right\} = {\gamma _1}{\left\| {{{{\bf{\bar h}}}_2}{\bf{\Phi }}{{{\bf{\bar H}}}_1}} \right\|^2} + {\gamma _2}MN + M.
		\end{equation}
	\end{IEEEproof}
	
	Under the high signal-to-noise ratio (SNR) condition, the upper bound derived for the ergodic SE in (\ref{eq:SE_ub}) is tight and can be regarded as a good approximation\cite{7727938}. Changing the arrangement of RIS from UPA to ULA, the upper bound is degenerated to the same result as in\cite{RN10}, which is a special case when RIS is arranged as a ULA.
	
	\textit{Lemma 1} shows that when the parameters related to hardware configuration and signal propagation environment such as $P$, ${{\bf{\bar h}}_2}$, ${{\bf{\bar H}}_1}$, ${K_1}$, ${K_2}$ are fixed, the upper bound ergodic SE is dependent on ${\| {{{{\bf{\bar h}}}_2}{\bf{\Phi }}{{{\bf{\bar H}}}_1}} \|^2}$, which is further determined by phase shift settings of RIS. By properly setting the phase shifter of the RIS to elevate the value of ${\| {{{{\bf{\bar h}}}_2}{\bf{\Phi }}{{{\bf{\bar H}}}_1}} \|^2}$, the upper bound of ergodic SE can be improved. Therefore, in the following, we devote to investigating the optimal design of ${\bf{\Phi }}$ to maximize the upper bound of ergodic SE.
	\subsection{Optimal Phase Shift Design}\label{sec:theory_B}
	To maximize the ergodic SE, we proposed an optimal design in the following lemma.
	
	\begin{lemma}
		Based on (\ref{eq:SE_ub}), when $P$, ${K_1}$, ${K_2}$, $M$, and $N$ remain unchanged, maximizing the tight upper bound ${C^{ub}}$ is equivalent to maximizing the value of ${\| {{{{\bf{\bar h}}}_2}{\bf{\Phi }}{{{\bf{\bar H}}}_1}} \|^2}$. The optimal ${\bf{\Phi }}$ is found as
		\begin{equation}
			{{\bf{\Phi }}_{\rm{opt}}} = \rm{blkdiag}\left\{ {{{\bf{\Phi }}_{1,\rm{opt}}},...,{{\bf{\Phi }}_{q,\rm{opt}}},...,{{\bf{\Phi }}_{Q,\rm{opt}}}} \right\},
		\end{equation}
		with ${{\bf{\Phi }}_{q,\rm{opt}}} = \rm{diag}({e^{j{\phi _{q,\rm{opt}}}}},...,{e^{j{\phi _{q,\rm{opt}}}}})$ and ${\phi _{q,\rm{opt}}}$ given by
		\begin{equation}
			{\phi _{q,\rm{opt}}} =  - [2{p_1}({x_q} - 1)   + 2{p_2}({y_q} - 1)   
			+ {p_1}({L_x} - 1)   + {p_2}({L_y} - 1)  ],
		\end{equation}
		where ${x_q}$ and ${y_q}$ are the horizontal and vertical coordinates of the first element of the $q$-th subarray in RIS, and ${L_x}$ and ${L_y}$ denote the row and column number of the $q$-th subarray in RIS. In addition, ${p_1}$ and ${p_2}$ are given by
		\begin{equation}
			\hspace{-0.8em}
			\begin{cases}
				{p_1} = \pi ({d_2}/{\lambda }) (\sin {\theta _{d_2}} - \sin {\theta _{a_1}})\\
				{p_2} = \pi ({d_2}/{\lambda }) (\sin {\varphi _{d_2}}\cos {\theta _{d_2}} - \sin {\varphi _{a_1}}\cos {\theta _{a_1}})
			\end{cases}
			\hspace{-0.8em}
			.
			\label{p_1,2}
		\end{equation}
	\end{lemma}
	
	\begin{IEEEproof}
		Please refer to Appendix \ref{B}.
	\end{IEEEproof}
	
	Utilizing \textit{Lemma 1} and \textit{2}, the maximum upper bound of the ergodic SE is obtained when the above optimal design is adopted, as shown in \textit{Theorem 1}.
	
	\begin{theorem}
		The upper bound ergodic SE of a subarray-based RIS-assisted system is maximized as
		\begin{equation}
			\begin{cases}
				C_{\rm{subarray}}^{\rm{ub}} = {\log _2}(1 + PM({\gamma _1}\eta {N^2} + {\gamma _2}N + 1))\\
				\eta  = {\left| {\frac{{\sin ({L_x}  {p_1})  \sin ({L_y}  {p_2})}}{{{L_x}\sin ({p_1})  {L_y}\sin ({p_2})}}} \right|^2} \in [0,1)
			\end{cases},
			\label{SE_pmax}
		\end{equation}
		where ${p_1}$ and ${p_2}$ are given in (\ref{p_1,2}).
	\end{theorem}
	
	\begin{IEEEproof}
		Please refer to Appendix \ref{C}.
	\end{IEEEproof}

	By comparing (\ref{eq:SE_ub}) and (\ref{SE_pmax}), we can conclude that the LoS component always contributes to the ergodic SE when the phase shift design is appropriate, while the degree to which it works is dictated by the value of $\eta$, which depends on the hardware configuration of RIS, including element spacing, arrival angle, departure angle, and the row and column number of the subarray. In addition, the ergodic SE increases with the number of RIS elements. By enlarging the size of RIS, the power of the reflected signal received is rising, which further yields an increase in ergodic SE.    
	
	\begin{corollary}
		When the subarray-based RIS degrades to element-based RIS, which can be considered as a subarray-based RIS where the subarray consists of only one element, the maximized upper bound of ergodic SE in an element-based RIS-assisted system can be given by
		\begin{equation}
			{{C}}_{\rm{element}}^{\rm{ub}}{\rm{ = }}{\log _2}(1 + PM({\gamma _1}{N^2} + {\gamma _2}N + 1)).
			\label{SE_emax}
		\end{equation}    	
	\end{corollary}

	\begin{IEEEproof}
		\textit{Corollary 1} can be shown by calculating terms in (\ref{SE_pmax}) as it is obvious in this special case that $L_x = L_y =1 $.
	\end{IEEEproof}

	In large-scale systems, namely, $M$ or $N$ 
	is large enough, the difference between the maximized upper bound of ergodic SE in a subarray-based RIS-assisted system and that in an element-based RIS-assisted system is approximated by
	\begin{equation}
		\begin{aligned}
			\Delta {C^{\rm{ub}}} =& {{C}}_{\rm{element}}^{\rm{ub}}-C_{\rm{subarray}}^{\rm{ub}}\\
			=& {\log _2}\left( {\frac{{1 + PM({\gamma _1}{N^2} + {\gamma _2}N + 1)}}{{1 + PM({\gamma _1}\eta {N^2} + {\gamma _2}N + 1)}}} \right)\\
			\approx& {\log _2}\left( {\frac{{PM({\gamma _1}{N^2} + {\gamma _2}N + 1)}}{{PM({\gamma _1}\eta {N^2} + {\gamma _2}N + 1)}}} \right)\\
			=& {\log _2}\left( {\frac{{{\gamma _1}{N^2} + {\gamma _2}N + 1}}{{{\gamma _1}\eta {N^2} + {\gamma _2}N + 1}}} \right).
		\end{aligned}
	\end{equation}
	In this case, $\Delta {C^{\rm{ub}}}$ is only related to the Rician factor and $\eta$.
		\vspace{-0.1cm}
	\subsection{Effects of Rician Factors and $\eta$}\label{sec:effects}
	 To acquire deep insights into the effects of Rician factors and $\eta$, $P$, $M$, and $N$ are assumed to be constant in this subsection. According to (\ref{SE_pmax}),  $C_{\rm{subarray}}^{\rm{ub}}$ is only determined by Rician factors and $\eta$.
	\subsubsection{effects of Rician factors}
	\subsubsection*{Case 1} When ${K_1}{\rm{ = }}0$ or ${K_2}{\rm{ = }}0$, by substituting the condition into (\ref{eq:SE_ub}) and (\ref{gamma}), 
	the upper bound of ergodic SE in both subarray-based and element-based RIS-assisted systems is calculated as
	\begin{equation}
		C_{\rm{subarray}}^{\rm{ub}}{\rm{ = }}{\log _2}(1 + PM(N + 1)).
	\end{equation}
	
	It can be shown that ergodic SE is proportional to $M$ and $N$ but is independent of the phase shift matrix ${\bf{\Phi }}$ due to the spatial isotropy of the auxiliary channel, which is insensitive to beamforming in channels ${{\bf{H}}_1}$ and ${{\bf{h}}_2}$. Thus, matrix ${\bf{\Phi }}$ can be set arbitrarily. Moreover, in such undesirable conditions, the performance of the upper bound of ergodic SE in a subarray-based RIS-assisted system is identical to that in an element-based RIS-assisted system. Thus, the subarray-based RIS is preferred in terms of EE due to its lower power consumption.
	
	\subsubsection*{Case 2} When ${K_1},{K_2} \to \infty $, then it is straightforward to have ${\gamma _1} \to 1,{\gamma _2} \to 0$, which refers to an extreme Rician fading environment where there exist only LoS components of channels. Utilizing (\ref{gamma}) and (\ref{SE_pmax}), the upper bound of ergodic SE in a subarray-based RIS-assisted system is maximized as
	\begin{equation}
		C_{\rm{subarray}}^{\rm{ub}} = {\log _2}(1 + PM(\eta {N^2} + 1)).
	\end{equation}
	
	In large-scale systems, with $N$ large enough, the difference between the maximized upper bound of ergodic SE in a  subarray-based RIS-assisted system and that in an element-based RIS-assisted system is given by
	\begin{equation}
		\Delta {C^{\rm{ub}}} 
		\overset{{\gamma _1} \to 1,{\gamma _2} \to 0}{=} {\log _2}\left( {\frac{{{N^2} + 1}}{{\eta {N^2} + 1}}} \right)
		\overset{N \to \infty}{=} - {\log _2}\left( \eta  \right).
		\label{eq:SE_dif-appr}
	\end{equation}

	It is worth mentioning that the difference above is only dependent on $\eta$, which further depends on the hardware configuration of RIS, including element spacing, arrival angle, departure angle, and the size of the subarray.
	\subsubsection{effects of $\eta$}\label{eta_effect}
	\subsubsection*{Case 1}
	When the signal is specularly reflected at RIS, which means ${\theta _{d_2}} = {\theta _{a_1}},{\varphi _{d_2}} = {\varphi _{a_1}}$, the maximized upper bound of ergodic SE in (\ref{SE_pmax}) is calculated as
	\begin{equation}
		C_{\rm{subarray}}^{\rm{ub}} = {\log _2}(1 + PM({\gamma _1}{N^2} + {\gamma _2}N + 1)).
		\label{SE_pmaxc1}
	\end{equation}
	
	\setlength{\topmargin}{-0.71in}
	
	Comparing (\ref{SE_pmaxc1}) with (\ref{SE_emax}), it is straightforward that the maximized upper bound of ergodic SE in a subarray-based RIS-assisted system equals to that in an element-based RIS-assisted system. Considering the power consumption reduction introduced by the subarray scheme of RIS in (\ref{EE_second}), it can be shown that when RIS happens to specularly reflect signals, the ergodic EE of a subarray-based RIS-assisted system is always higher than that of an element-based RIS-assisted system.
	\subsubsection*{Case 2} 
	When ${L_x}{p_1} = k\pi $ or ${L_y}{p_2} = k\pi $, the maximized upper bound of ergodic SE in (\ref{SE_pmax}) is rewritten as
	\begin{equation}
		C_{\rm{subarray}}^{\rm{ub}} = {\log _2}(1 + PM({\gamma _2}N + 1))
	\end{equation}
	
	This is the most disadvantageous case in which the upper bound of ergodic SE in subarray-based RIS-assisted systems takes its minimum. The explanation for this phenomenon is that the condition ${L_x}{p_1} = k\pi $ or ${L_y}{p_2} = k\pi $ means a coincidence of the user position and the zero point of the reflected beam on each subarray, which disenables the wireless signal beam to form on the main lobe of the assistant channel. In this case, RIS can only work as a reflecting surface in the scattering path of the signal.
	
	In this case, or more generally, it can be derived from (\ref{SE_pmax}) and (\ref{SE_emax}) that when the RIS hardware configuration (i.e. subarray setting) is fixed, the upper bound of ergodic SE of a subarray-based RIS-assisted system is always lower or equal to that of an element-based RIS-assisted system where the equality relation holds only in \textit{Case 1} of Section \ref{eta_effect}, and the difference between them increases logarithmically as the number of RIS units increases. Meantime, according to (\ref{P_RIS}) and (\ref{P_total drive circuits}), the total power consumption of the subarray-based RIS-assisted system is always smaller than that of the element-based RIS-assisted system, and the difference between them increases linearly as the number of RIS units increases. Therefore, with the increase in the number of RIS elements, the maximized upper bound of ergodic SE in the element-based RIS-assisted system over that of the subarray-based RIS-assisted system tends to grow more slowly, while the excess of total power consumption is still linearly increasing.
	
	As stated above, it can be concluded theoretically that when the number of RIS elements is large enough, the ergodic EE of a subarray-based RIS-assisted system is higher than that of an element-based RIS-assisted system. Considering the multiplicity of factors that contributes to the value of ergodic EE and the difficulty of solving the analytic solutions associated with logarithmic functions, quantitative analysis on ergodic EE can become very challenging. In the next section, the ergodic EE of subarray-based and element-based RIS-assisted systems is compared and analyzed by numerical results instead.
	\section{Numerical Results}\label{sections:Numerical_result}
	In this section, we numerically verified the tightness of the maximized upper bound derived in Section \ref{sec:theory_B}, then the effects of the subarray scheme on ergodic SE and EE are also evaluated by comparing with that of the element-based RIS-assisted system. We set the number of BS antennas and RIS elements as $M = 64$ and $N = {N_x} \times {N_y} = 32 \times 32 = 1024$, respectively. The vertical and horizontal spacing between BS adjacent antennas and RIS adjacent elements are set to half carrier wavelength. The transmission signal-to-noise ratio is 10 dB. Without loss of generality, ${K_1}$ and ${K_2}$ are considered the same and denoted by $K$. Note that the optimal phase shift design is adopted in all systems.
	
	\vspace{-0.3cm}
	\begin{figure}[htbp]
	\centering
	\includegraphics[width=0.4\textwidth]{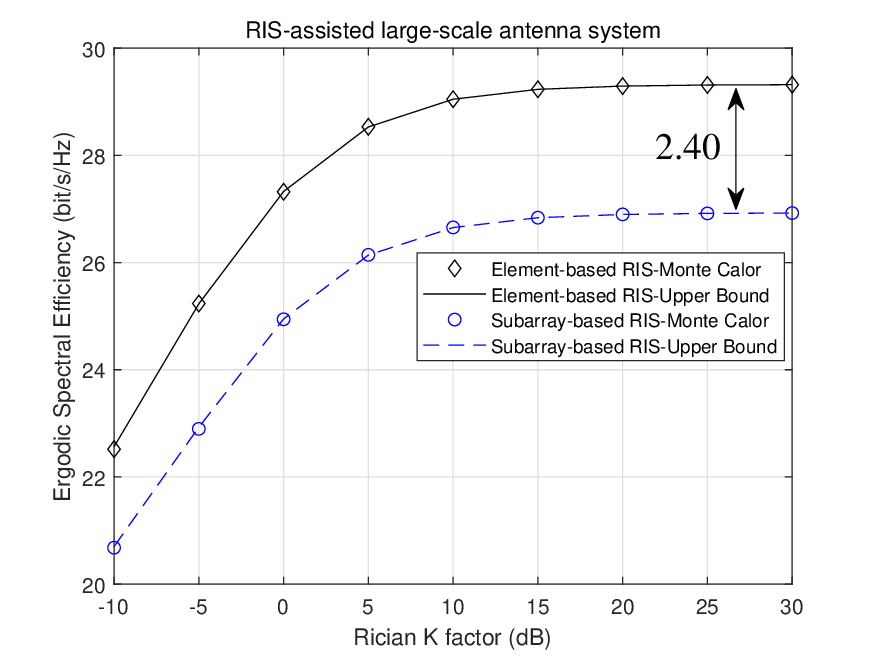}
	\caption{. Comparison of the Monte Carlo results and ergodic $SE$}
	\label{fig:testk}
    \end{figure}
	\vspace{+0.3cm}
	
	The tightness of the maximized upper bound is illustrated in Fig. \ref{fig:testk} which shows ergodic SE versus different values of the Rician factor. Here we set ${L_x} = {L_y} = {L_0} = 2$ representing the size of the subarray, and ${\theta _{d_1}} = \frac{1}{2}\pi ,{\varphi _{a_1}} = \frac{7}{6}\pi ,{\theta _{a_1}} = \frac{2}{3}\pi ,{\varphi _{d_2}} = \frac{4}{3}\pi ,{\theta _{d_2}} = \frac{5}{3}\pi $ denoting angles of arrival and departure at BS and RIS respectively. In this setting, $\eta =0.19$ and $- {\log _2}\left( \eta  \right) =2.40$. It is shown that the maximized upper bound is highly approximated to the Monte Carlo results and the gap between them diminishes with the increase of the Rician factor. When the Rician factor increases, the maximized upper bounds of ergodic SE in a subarray-based RIS-assisted system and in an element-based RIS-assisted system approach to constants with a difference being 2.40, which is consistent with the results calculated based on (\ref{eq:SE_dif-appr}).
	
	\vspace{-0.3cm}
	\begin{figure}[htbp]
	\centering
	\includegraphics[width=0.4\textwidth]{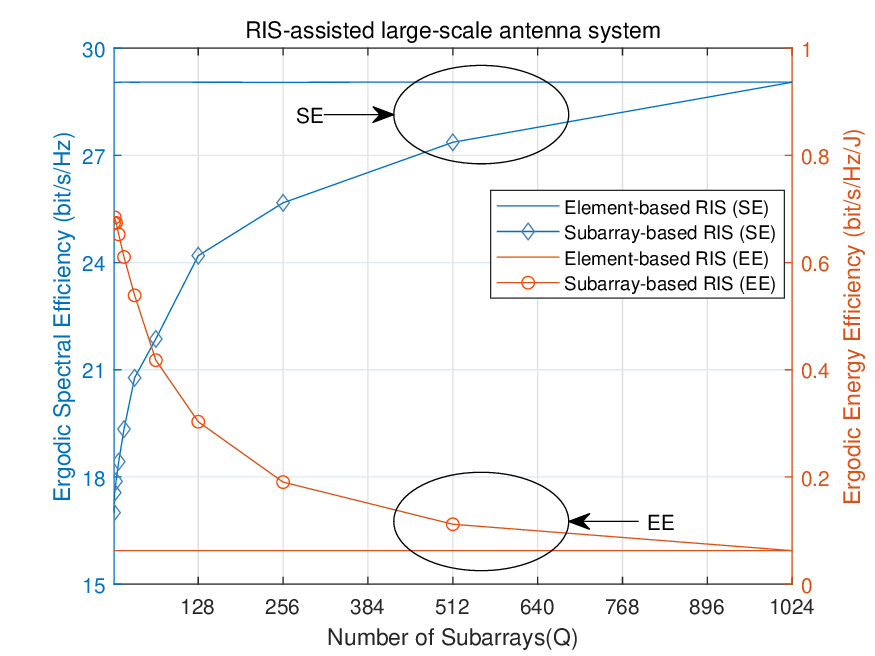}
	\caption{. Regional ergodic SE and EE performance versus $Q$}
	\label{fig:testQ}
    \end{figure}	
	\vspace{+0.3cm}
	
	\setlength{\topmargin}{-0.7in}
	
	To obtain a continuously adjustable phase for RIS, a varactor-diode-based RIS is adopted, with its relevant power-consumption parameters set referring to\cite{2023arXiv230300299W} as follows: ${P_{{\rm{dy}}}} \approx 0,{P_{{\rm{ct}}}} = 4.8{\rm{ W}}$ and ${P_{{\rm{d}}}} = 430{\rm{ mW}}$ denote the dynamic power consumption of RIS units, the power consumption of the FPGA board, and the rated power consumption of a single drive circuit, respectively. The power consumption of the system exclusive of RIS is set as ${P_{{\rm{rest}}}} = 20{\rm{ W}}$. Without loss of generality, the number of drive circuits ${N_{{\rm{d}}}}$ is set to equal to the number of subarrays $Q$. To evaluate the effects of related parameters on the regional SE and EE of the system, the arrival and departure angles in the system are set to random values. Fig. 3 shows the regional SE and EE versus the different numbers of the subarrays. It is observed that as the number varies from $1$ to $1024$, when a subarray scheme is adopted there are levels of loss in the regional SE. However, there has been a corresponding increase in regional EE. This is because the number of RIS units and the power consumption of the drive circuit is fairly large, which contributes to the high proportion of RIS power consumption in the total power consumption of the system. Therefore, the increase of ergodic EE on account of the power consumption reduction introduced by the subarray scheme exceeds the decrease of ergodic EE caused by the reduction of ergodic SE. which gives a reasonable explanation for its increase.
	
	\vspace{-0.3cm}
    \begin{figure}[htbp]
		\centering
		\includegraphics[width=0.4\textwidth]{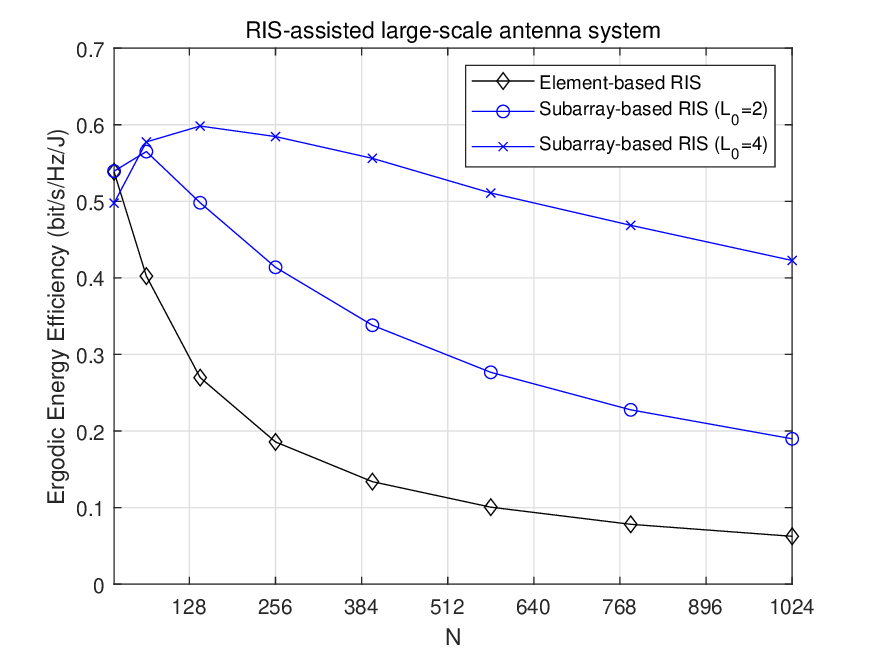}
		\caption{. Regional ergodic EE performance versus $N$}
		\label{fig:testN}
	\end{figure}
	\vspace{+0.3cm}
	
	In Fig. \ref{fig:testN}, we evaluate the regional ergodic EE versus the scale of RIS, which is usually large in practical deployment to achieve higher reflected power. We consider ${L_x}$ and ${L_y}$ as the same and denote them by ${L_0}$, with the value of ${L_0}$ in the set of  $\{ 2,4\} $. As discussed in Section \ref{sec:effects}, with the increase in the number of RIS units, the growth of the maximized upper bound of the ergodic SE is slowing down while the total power consumption keeps linearly increasing. As shown in Fig. \ref{fig:testN}, when $N$ is small, the reduction of ${P_{\rm{tot}}}$ introduced by the subarray-based scheme disables to compensate for the loss of ergodic EE caused by the decrease of ergodic SE, which further leads to the result that the ergodic EE of the subarray-based RIS-assisted system is lower than that of the element-based RIS-assisted system. In contrast, when $N$ increases, the ergodic EE of the  subarray-based RIS-assisted system is higher than that of the element-based RIS-assisted system due to the expanded advantages of the subarray-based RIS in reducing power consumption.

	\section{Conclusion}\label{section:conclusion}
	In this study, the effects of the subarray-based RIS on the ergodic SE and EE are investigated. To gain a direct perception of the ergodic SE, we first derived an upper bound for the ergodic SE. Under the subarray-based scheme, we proposed an optimal phase shift design to maximize the upper bound. Furthermore, we analyzed the gap between ergodic SE as well as EE in the subarray-based scheme and in the element-based scheme based on the analytical expression. Numerical results verified the tightness of the upper bound as well as the effectiveness of the proposed optimal phase shift design. The dependence of the ergodic SE and EE on the element numbers and the subarray size is also revealed in numerical results.

	\appendices

	\section{}\label{B}
	For fixed $P$, ${K_1}$, ${K_2}$, $M$, and $N$, the problem equals to solving
	\begin{equation}
		{{\bf{\Phi }}_{\rm{opt}}} = \mathop {\max }\limits_{\bf{\Phi }} {\left\| {{{{\bf{\bar h}}}_2}{\bf{\Phi }}{{{\bf{\bar H}}}_1}} \right\|^2},
		\label{eq:prob}
	\end{equation}
    where ${\| {{{{\bf{\bar h}}}_2}{\bf{\Phi }}{{{\bf{\bar H}}}_1}} \|^2}$ can be calculated as in (\ref{eq:calh}).

	\begin{figure*}[!htb]
		\centering
		\begin{equation}
			\begin{aligned}
				{{{\bf{\bar h}}}_2}{\bf{\Phi }}&{{{\bf{\bar H}}}_1}=[{\bf{c}} \otimes {{\bf{a}}_L}({\theta _{d_2}},{\varphi _{d_2}})]
				{\rm{blkdiag}}\left[ {\begin{array}{*{20}{c}}
						{{{\bf{\Phi }}_1}}&{...}&{{{\bf{\Phi }}_q}}&{...}&{{{\bf{\Phi }}_Q}}
				\end{array}} \right]
				\{{\bf{ b}}^T\otimes[{\bf{a}}_L^H({\theta _{a_1}},{\varphi _{a_1}}){{\bf{a}}_M}({\theta _{d_1}})]\}\\				
				=&\left[ {\begin{array}{*{20}{c}}
						{{c_1}{{\bf{a}}_L}({\theta _{d_2}},{\varphi _{d_2}}){{\bf{\Phi }}_1}}&{...}&{{c_Q}{{\bf{a}}_L}({\theta _{d_2}},{\varphi _{d_2}}){{\bf{\Phi }}_Q}}
				\end{array}} \right]
				\left[ {\begin{array}{*{20}{c}}
						{{b_1}{{\bf{a}}^H_L}({\theta _{a_1}},{\varphi _{a_1}}){{\bf{a}}_M({\theta_{d_1}})}}&{...}&{{b_Q}{{\bf{a}}^H_L}({\theta _{a_1}},{\varphi _{a_1}}){{\bf{a}}_M({\theta_{d_1}})}}
				\end{array}} \right]^T\\		    
				=&\sum\nolimits_q{{c_q}{b_q}{{{\bf{a}}_L}({\theta _{d_2}},{\varphi _{d_2}})}{{\bf{\Phi }}_q}{{\bf{a}}_L^H({\theta _{a_1}},{\varphi _{a_1}})}{{\bf{a}}_M({\theta_{d_1}})}}.
			\end{aligned}
			\label{eq:calh}
		\end{equation}
		\vspace{-0.6cm}
		{\noindent}\rule[0pt]{18cm}{0.05em}
	\end{figure*}
	
	By exploiting (\ref{eq:calh}), (\ref{eq:prob}) can be modified as
	\begin{equation}
		{{\bf{\Phi }}_{\rm{opt}}} = \mathop {\max }\limits_{\bf{\Phi }} {\left\| {\sum\nolimits_q {{z_q}{{\bf{a}}_M}({\theta _{d_1}})} } \right\|^2}\\
		= \mathop {\max }\limits_{\bf{\Phi }} {\left| {\sum\nolimits_q {{z_q}} } \right|^2}{\left\| {{{\bf{a}}_M}({\theta _{d_1}})} \right\|^2},
		\label{eq:optphi}
	\end{equation}
	where ${\left\| {{{\bf{a}}_M}({\theta _{d_1}})} \right\|^2} = M$ is a constant and ${z_q}$ is defined as  ${z_q}={c_q}{b_q}{{{\bf{a}}_L}({\theta _{d_2}},{\varphi _{d_2}})}{{\bf{\Phi }}_q}{{\bf{a}}_L^H({\theta _{a_1}},{\varphi _{a_1}})}$. Therefore, optimizing ${\bf{\Phi }}$ equals to maximizing ${| {\sum\nolimits_q {{z_q}} } |^2}$. To solve the optimization problem above, ${z_q}$ can be calculated as 
	\begin{equation}
		{z_q} = e^{j({\phi _q} + {\phi _1} + {\phi _2})}{z'_q},
	\end{equation}
	where
	\begin{equation}
		\begin{aligned}
			&\phi _1=2\pi ({d_2}/{\lambda })[(\sin {\theta _{d2}} - \sin {\theta _{a1}})({x_q} - 1)\\
			 + (\sin& {\varphi _{d2}}\cos {\theta _{d2}} - \sin {\varphi _{a1}}\cos {\theta _{a1}})({y_q} - 1)],
		\end{aligned}
	\end{equation}
    \begin{equation}
    	\phi _2=[({L_x} - 1)  {p_1} + ({L_y} - 1)  {p_2}],
    \end{equation}
    \begin{equation}
    	z'_q=[{{\sin ({L_x}{p_1})\sin ({L_y}{p_2})}}]/[{{\sin ({p_1})\sin ({p_2})}}].
    \end{equation}

	Thus, we can obtain
	\begin{equation}
		\sum\nolimits_q {{z_q}}  = \sum\nolimits_q {{{e^{j({\phi _q} + {\phi _1} + {\phi _2})}}}{{z}'_q}} ,
		\label{eq:sumzq}
	\end{equation}
	where ${z'}_q$ remains unchanged for each subarray. Thus, the optimal phase shift is given by
	\begin{equation}
		{\phi _{q,\rm{opt}}} = {\phi _1} + {\phi _2}  .
		\label{eq:optelem}	
	\end{equation}

	\section{}\label{C}
	When the optimal phase shift design above is adopted, by exploiting (\ref{eq:optphi}), (\ref{eq:sumzq}), and (\ref{eq:optelem}), we have
	
	\begin{equation}
		{\left| {\sum\nolimits_q {{z_q}} } \right|^2} = {\left| {Q {{z}'_q}} \right|^2}
		= {\left| {\frac{{\sin ({L_x}  {p_1})  \sin ({L_y}  {p_2})}}{{{L_x}\sin ({p_1})  {L_y}\sin ({p_2})}}} \right|^2}{N^2}
		=\eta {N^2}.
		\label{eq:optsumzq}
	\end{equation}

	Substituting (\ref{eq:optphi}) and (\ref{eq:optsumzq}) into (\ref{eq:SE_ub}), the maximized upper bound ergodic spectral efficiency is calculated as
	\begin{equation}
		C_{\rm{subarray}}^{\rm{ub}} =
		{\log _2}(1 + PM({\gamma _1}\eta {N^2} + {\gamma _2}N + 1)).
	\end{equation}
	
	Finally, (\ref{SE_pmax}) is obtained.
	
	\bibliographystyle{IEEEtran}
	\bibliography{IEEEabrv,reference}

\end{document}